\documentclass[twocolumn]{aastex62}

\usepackage{xcolor} 
\definecolor{bg-green}{rgb}{0.8588,0.9333,0.8666}
\definecolor{Q-color}{rgb}{0.5,0.1,0.5}

\usepackage{CJK}
\usepackage{epsf, graphicx, subfigure}
\usepackage{amsmath} 
\usepackage{natbib}
\usepackage{color}
\usepackage{colortbl}
\usepackage{hyperref}
\hypersetup{
    colorlinks,%
    citecolor=blue,%
    filecolor=black,%
    linkcolor=blue,%
    urlcolor=black
}

\renewcommand{\arraystretch}{1.2}

\shorttitle{Using H$\alpha$ Filaments to Probe AGN Feedback}
\shortauthors{Qiu et al.}

\begin{document} 

\begin{CJK}{UTF8}{bsmi} 


\title{Using H$\alpha$ Filaments to Probe AGN Feedback in Galaxy Clusters}

\author{Yu Qiu (邱宇)$^\dagger$}
\email{$^\dagger$geoyuqiu@gatech.edu}
\affiliation{Center for Relativistic Astrophysics, Georgia Institute of Technology, Atlanta, GA 30332}

\author{Tamara Bogdanovi\'c$^\ddagger$}
\email{$^\ddagger$tamarab@gatech.edu}
\affiliation{Center for Relativistic Astrophysics, Georgia Institute of Technology, Atlanta, GA 30332}

\author{Yuan Li}
\affiliation{Department of Astronomy, University of California, Berkeley, CA 94720}

\author{Michael McDonald}
\affiliation{Kavli Institute for Astrophysics and Space Research, MIT, Cambridge, MA 02139}


\begin{abstract}
Recent observations of giant ellipticals and brightest cluster galaxies (BCGs) provide tentative evidence for a correlation between the luminosity of the H$\alpha$ emitting gas filaments and the strength of feedback associated with the active galactic nucleus (AGN). Motivated by this, we use 3D radiation-hydrodynamic simulations with the code \texttt{Enzo} to examine and quantify the relationship between the observable properties of the H$\alpha$ filaments and the kinetic and radiative feedback from supermassive black holes in BCGs. We find that the spatial extent and total mass of the filaments show positive correlations with AGN feedback power and can therefore be used as probes of the AGN activity. We also examine the relationship between the AGN feedback power and velocity dispersion of the H$\alpha$ filaments and find that the kinetic luminosity shows a statistically significant correlation with the component of the velocity dispersion along the jet axis, but not the components perpendicular to it.
\end{abstract}


\section{Introduction} \label{sec:intro}
Observational evidence indicates that feedback from active galactic nuclei (AGNs), powered by accretion onto the supermassive black holes (SMBHs), has a profound impact on their host galaxies and clusters. In cool-core clusters (CCCs), AGN feedback is thought to be the primary mechanism which prevents catastrophic radiative cooling of the intracluster medium (ICM) on time scales below 1\,Gyr \citep{Fabian2012,McNamara2012}.

The cores of CCCs are however not devoid of cold gas. A study of 16 CCCs has revealed $10^9 - 10^{11.5}\,M_\odot$ of molecular gas within a $\sim{\rm few}\times10\,{\rm kpc}$ of their BCGs \citep{Edge2001}. Similarly, a substantial fraction of CCCs host partially ionized filamentary gas with temperature $T\sim 10^4\,$K, that is emitting copious amounts of H$\alpha$ photons \citep{Heckman1989, Hatch2007,McDonald2010,McDonald2011}.
 The H$\alpha$ filaments are thought to form out of marginally unstable ICM \citep{McDonald2010,McCourt2012, Sharma2012}, possibly triggered by AGN feedback \citep{Werner2014, Li2014a, Li2015, Prasad2015, Gaspari2012, Gaspari2015, Tremblay2015, McNamara2016, Voit2017, Hogan2017, Martizzi2019}. This hypothesis is supported by observations, which show a {\it positive} correlation between the amount of molecular gas and the jet power in early-type galaxies \citep{Babyk2018}. Along similar lines, \citet{Lakhchaura2018} find tentative evidence for a weak positive correlation between the AGN jet power and ${\rm H}\alpha+[\textsc{N\,ii}]$ luminosity in giant ellipticals. 

These findings imply that the presence of cold gas necessitates AGN feedback, but do not uniquely answer the question how the cold gas is affected by it. In a recent study based on 3D radiation-hydrodynamic simulations, we find that AGN feedback in CCCs promotes the formation of the H$\alpha$ filaments, and their presence is a signpost for an ongoing or recent outburst of AGN feedback \citep[][hereafter Q18]{Qiu2018}. In this work, we examine the relationship between the properties of the H$\alpha$ filaments and AGN feedback power. 


\section{Simulations of AGN Feedback} \label{sec:sim}

The suite of 3D radiation-hydrodynamic simulations analyzed in this work are carried out with the code \texttt{Enzo} \citep{Enzo2014}. Their key aspects are summarized below, and more detailed description of the numerical setup can be found in Q18. In this work we show results of the two representative high-resolution runs (RT02 and TI07). The main difference between them is in the modeling of radiation transport: in RT02 we calculate it explicitly using the ray-tracing module \textsc{moray}  \citep{Wise2011}, which accounts for photoionization, Compton heating, and radiation pressure associated with these processes. In TI07 we approximate it by injecting in the vicinity of the SMBH thermal energy commensurate to the energy of the radiation emitted by the central AGN. In all other regards the two runs are identical and produce consistent results.

The cluster is centered on a computation domain with size $(500\,{\rm kpc})^3$, and modeled with maximum numerical resolution of 0.24\,kpc at the core. The gravitational potential of the cluster is modeled as the sum of the NFW dark matter profile \citep{Navarro1996}, the BCG stellar bulge \citep{Mathews2006}, and the central SMBH \citep{Wilman2005}. Initial conditions for the ICM are based on the observed temperature and density profiles of the Perseus cluster \citep{Churazov2004, Mathews2006, Li2012}. The gas comprises 6 species: $\textsc{H\,i}$, $\textsc{H\,ii}$, $\text{He}\,\textsc{i}$, $\text{He}\,\textsc{ii}$, $\text{He}\,\textsc{iii}$, and $e^-$. The radiative cooling utilizes the non-equilibrium cooling implemented in \texttt{Enzo} for $\text{H}$ and $\text{He}$ species, supplemented by a cooling table to describe additional radiation losses due to metals \citep{Smith2008}. A constant value of metallicity, $Z=0.011$, is assumed throughout the cluster. We do not model molecular gas (or radiation pressure on dust).

Simulated AGN feedback is powered by accretion onto the central SMBH, such that the total feedback luminosity, $L= \eta \dot{M}_{\rm BH}c^2$, where $\eta=0.1$ is the feedback efficiency. $\dot{M}_{\rm BH}$ is dominated by accretion of the cold gas ($T<3\times 10^4 \,{\rm K}$),  and is estimated as $\approx \epsilon_{\rm acc}\,M_{\rm cg}/\tau$, where $\epsilon_{\rm acc} = 10^{-2}$  is the accretion efficiency, $M_{\rm cg}$ is the mass of the cold gas within the central 1\,kpc, and $\tau = 5$\,Myr is the free-fall timescale. The feedback power is allocated to the kinetic and radiative luminosity of the AGN as $L = L_{\rm K} + L_{\rm R}$. In this model, the AGN feedback is assumed to be radiatively inefficient when $\dot{M}_{\rm BH}\leq0.05\, \dot{M}_\text{Edd}$, where $L_{\rm R} \propto \dot{M}_{\rm BH}^2\leq L_{\rm K}$ (equality is reached at $0.05\, \dot{M}_\text{Edd}$). When $\dot{M}_{\rm BH} > 0.05\, \dot{M}_\text{Edd}$, the AGN transitions to the radiatively efficient regime, where its luminosity is equally divided between the kinetic and radiative modes, so that $L_\text{K}=L_\text{R}\propto\dot{M}_{\rm BH}$.


\section{H$\alpha$ Filaments as a Measure of AGN Feedback Power}  \label{sec:result} \label{sec:fila}

\begin{figure*}[t!]
\centering
\includegraphics[height=0.425\linewidth]{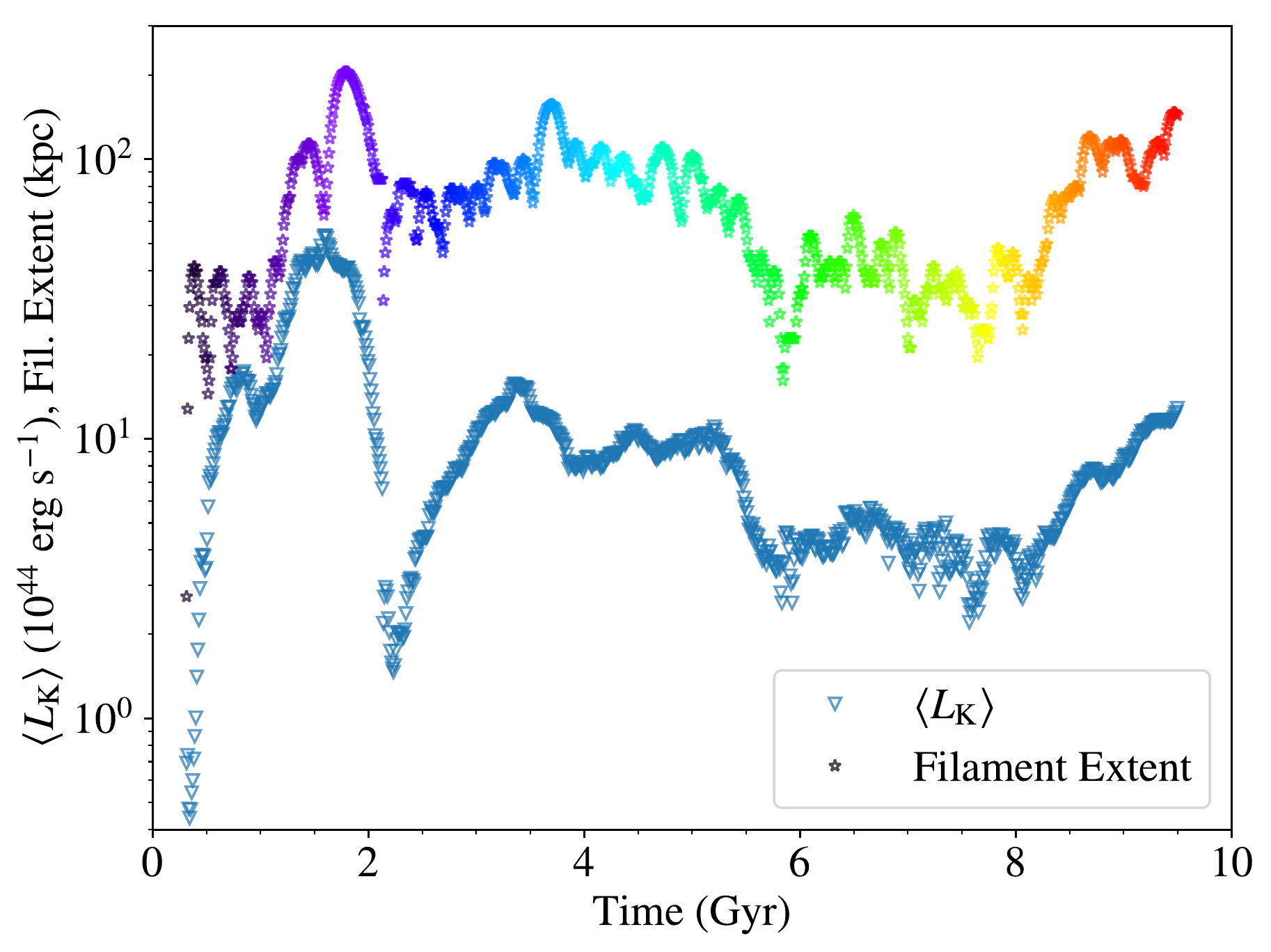}\includegraphics[height=0.425\linewidth]{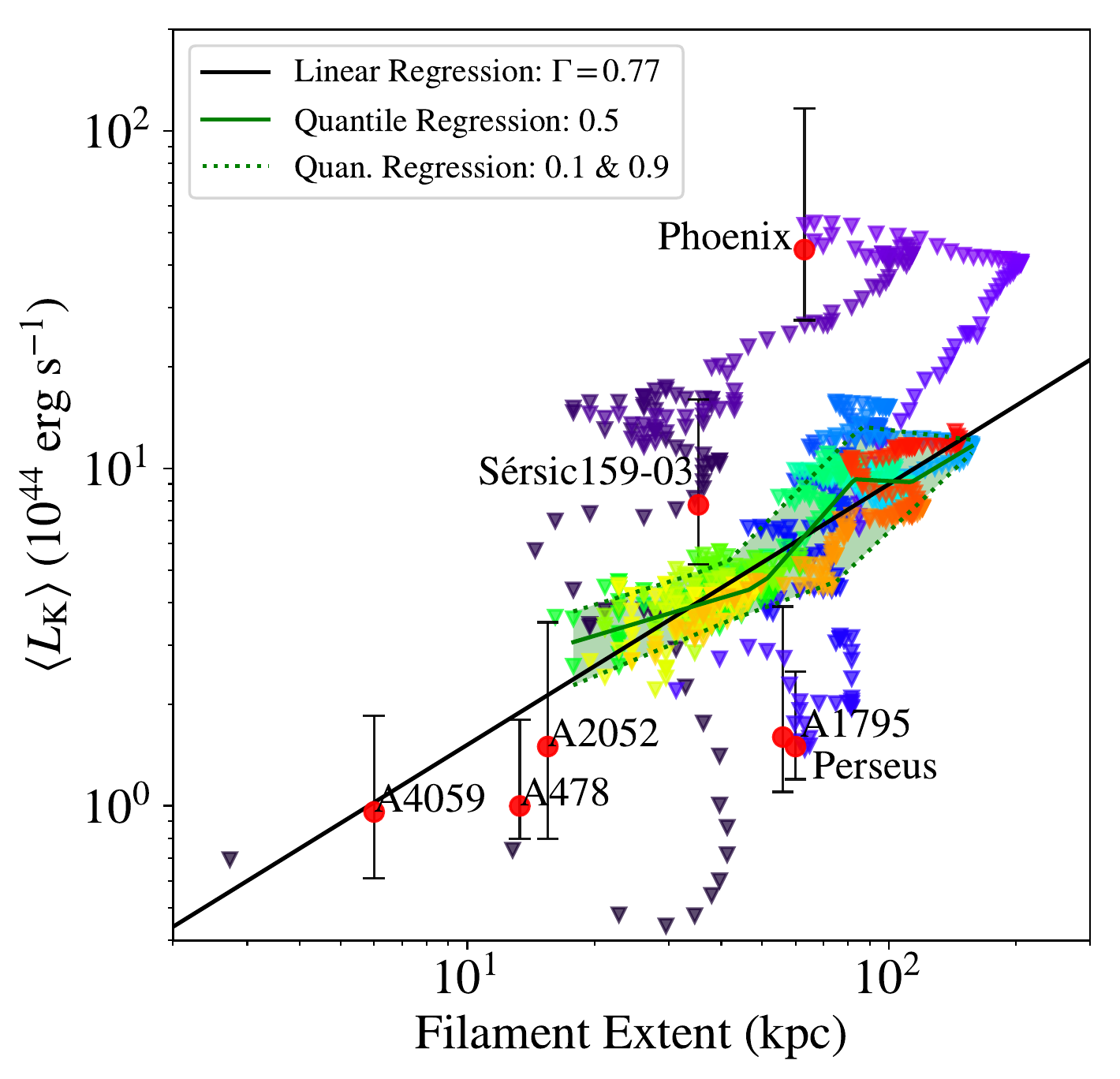}
\caption{{\it Left:} Evolution of the average kinetic luminosity, $\langle L_{\rm K}\rangle$ (blue triangles), and spatial extent of the filaments (multicolor stars) measured in simulation TI07. {\it Right:} $\langle L_{\rm K}\rangle$ as a function of filament extent. Colors mark different times in the simulation. The black solid line shows the power-law fit for data points between $2-9.5\,{\rm Gyr}$, while green lines indicate the 0.1, 0.5, and 0.9 quantiles for additive quantile regression. Red circles mark the observed clusters with extended H$\alpha$ filaments from \citet{McDonald2010, McDonald2015}, with additional X-ray cavity power and associated uncertainties from \citet{Rafferty2006}.}
\label{fig:f_extent}
\end{figure*}

In this section we consider several properties of the H$\alpha$ filaments measured from our simulations (spatial extent, mass, and velocity dispersion) and examine their correlations with AGN feedback power. Precisely how the filaments form (i.e., whether they form in situ or are launched from the center of the cluster) will be examined in a companion paper.

\subsection{Spatial Extent of the Filaments} \label{sec:fila_extent}

In this section we examine whether the radial extent of the filaments from the cluster center correlates with AGN luminosity. In order to determine the radial extent at a given time, $r_{\rm fil}$, we divide the simulated cluster into radial shells of thickness 1.7\,kpc, and identify the radius of the outermost shell in which the cold gas\footnote{In simulations, the cold gas is represented by $\textsc{H\,i}$. In reality, this gas phase also includes H$_2$, which can power intra-filament star formation.} mass exceeds $10^3\,M_\Sun$. We have verified that other values between $10^3-10^7\,M_\Sun$ give similar results. 

The left panel of Figure~\ref{fig:f_extent} shows the evolution of $r_{\rm fil}$ in simulation TI07, along with the kinetic luminosity associated with the AGN jets, averaged over the dynamical timescale of the filaments. This timescale is calculated as $\Delta t_{\rm fil} = r_{\rm fil}/v_{\rm out}$, where $v_{\rm out}\approx 500\,{\rm km\,s}^{-1}$ is the characteristic outflow speed of the filaments in simulations (e.g., $\Delta t_{\rm fil} \approx 10^8\,$yr for $r_{\rm fil} = 50\,$kpc). The average kinetic luminosity calculated in this way is similar to the luminosity inferred from the X-ray measurements of cavities inflated in the ICM by jets.
Visual inspection reveals that more powerful outbursts of AGN feedback tend to produce more spatially extended filaments.

The right panel of Figure~\ref{fig:f_extent} shows the kinetic luminosity as a function of the spatial extent of the filaments over time. Two distinct trends are visible: one pertains to the early time in the evolution ($t<2\,{\rm Gyr}$), when the H$\alpha$ filaments extend radially out in all directions from the cluster center, as illustrated in Figure~\ref{fig:f_vel}.  After $2\,{\rm Gyr}$, the subsequent AGN outbursts result in filaments that are more collimated along the jet axis, and the jet power and the extent of the filaments can be approximately represented by the linear regression fit:
\begin{equation}
	\langle L_{\rm K}\rangle= (9.0\pm0.1)\times10^{44}\,{\rm erg\,s}^{-1} \left(\frac{r_{\rm fil}}{\rm 100\,kpc}\right)^{0.77\pm0.02}\,\,.
\label{eq1}	
\end{equation}
Note however that the $\langle L_{\rm K}\rangle - r_{\rm fil}$ correlation is not strictly linear and is more accurately described by the additive quantile regression \citep{quantreg}, illustrated by the solid green line.

This correlation can be understood by considering the freely expanding filaments in a spherically symmetric gravitational potential, defined by the radial acceleration $g(r)$. In the absence of non-gravitational forces, the terminal radius ($r_f$) of filaments launched with initial velocity $v_ i$ from radius $r_i$ is determined by the conservation of energy
\begin{equation}
	\int_{r_i}^{r_f}g(r)dr = -\frac{1}{2}v_i^2 \,\,.
\label{eq2}
\end{equation}
Assuming that in the spatial region of interest, the acceleration can be represented by a power law, $g(r)\propto -r^\alpha$, then the specific kinetic energy at launch is $k_i=\frac{1}{2}v_ i^2 \propto r_f^{\alpha+1}$ (if $\alpha \neq -1$). In the scenario where the initial kinetic energy of the filaments is supplied by AGN feedback, this implies $L_{\rm K} \propto r_ f^{\alpha+1}$. In our simulations, $\alpha$ is defined by the cluster potential and varies from $-0.6$ to $-0.2$ between $10$ and $100\,{\rm kpc}$ (see Appendix~A in Q18). For the H$\alpha$ filaments that expand freely to $\sim$100\,kpc, this implies $L_{\rm K} \propto r_f^{0.8}$. Because the ICM is orders of magnitude more dilute than the filaments, the hydrodynamic forces   (i.e., ram pressure) acting on the filaments can be neglected. Therefore, a simple ballistic model presented here provides a satisfactory description of their dynamics.

The derived dependence closely corresponds to that shown in equation~\ref{eq1}, measured for the evolution of filaments after 2\,Gyr, indicating that in this stage the filaments that form in jet-driven outflows expand freely into the ICM. In contrast, during the first 2\,Gyr the expanding filaments collide and are scattered in all directions by the infalling filaments. As a consequence, their spatial extent is suppressed relative to the later stages of evolution, when the ``channel" is cleared for the filaments to expand along the jet axis. Thus, the spatial extent of the H$\alpha$ filaments also provides a direct measure of how far the outflows reach.

\begin{figure*}[t!]
\centering
\includegraphics[height=0.413\linewidth]{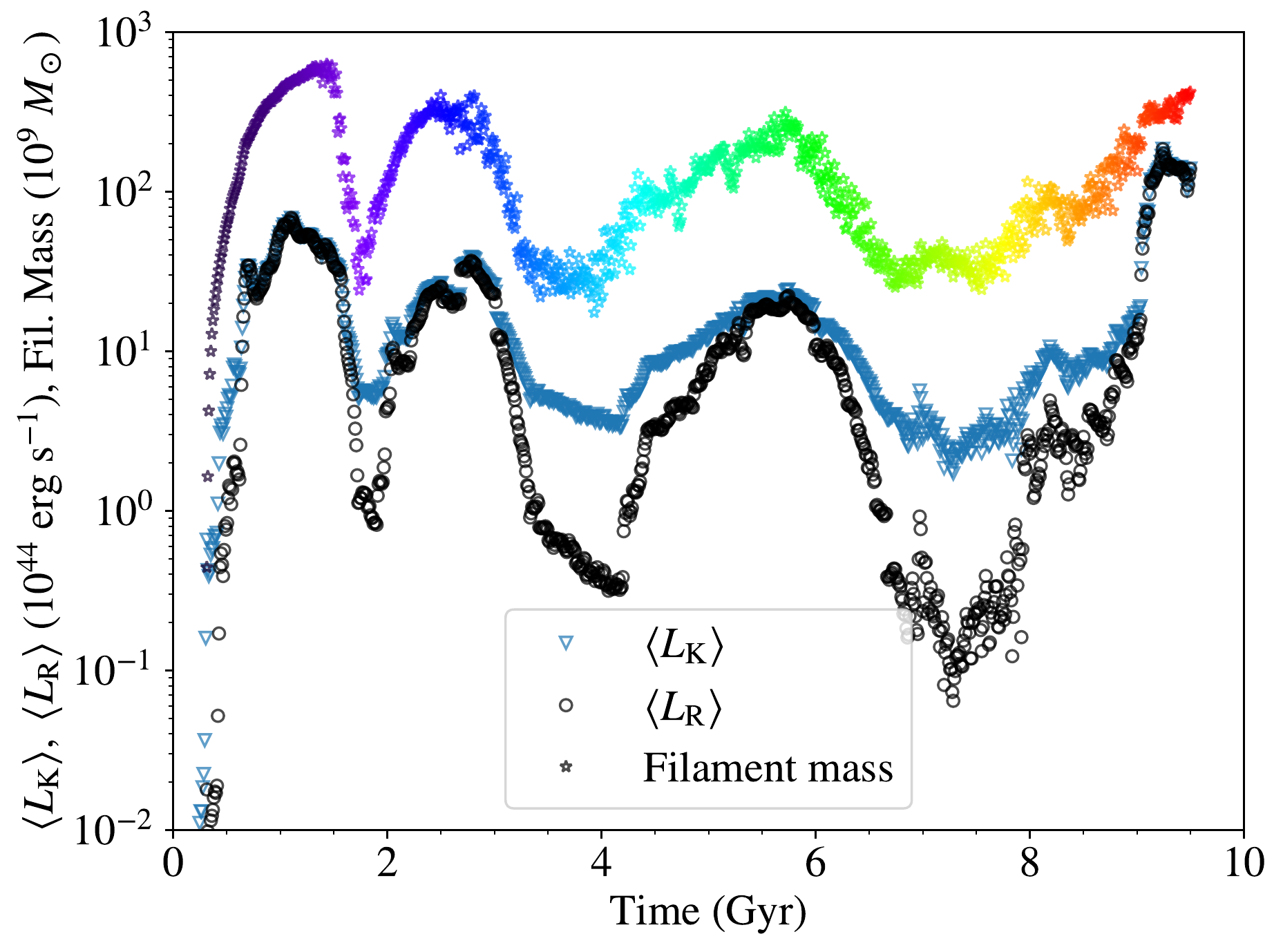}\includegraphics[height=0.413\linewidth]{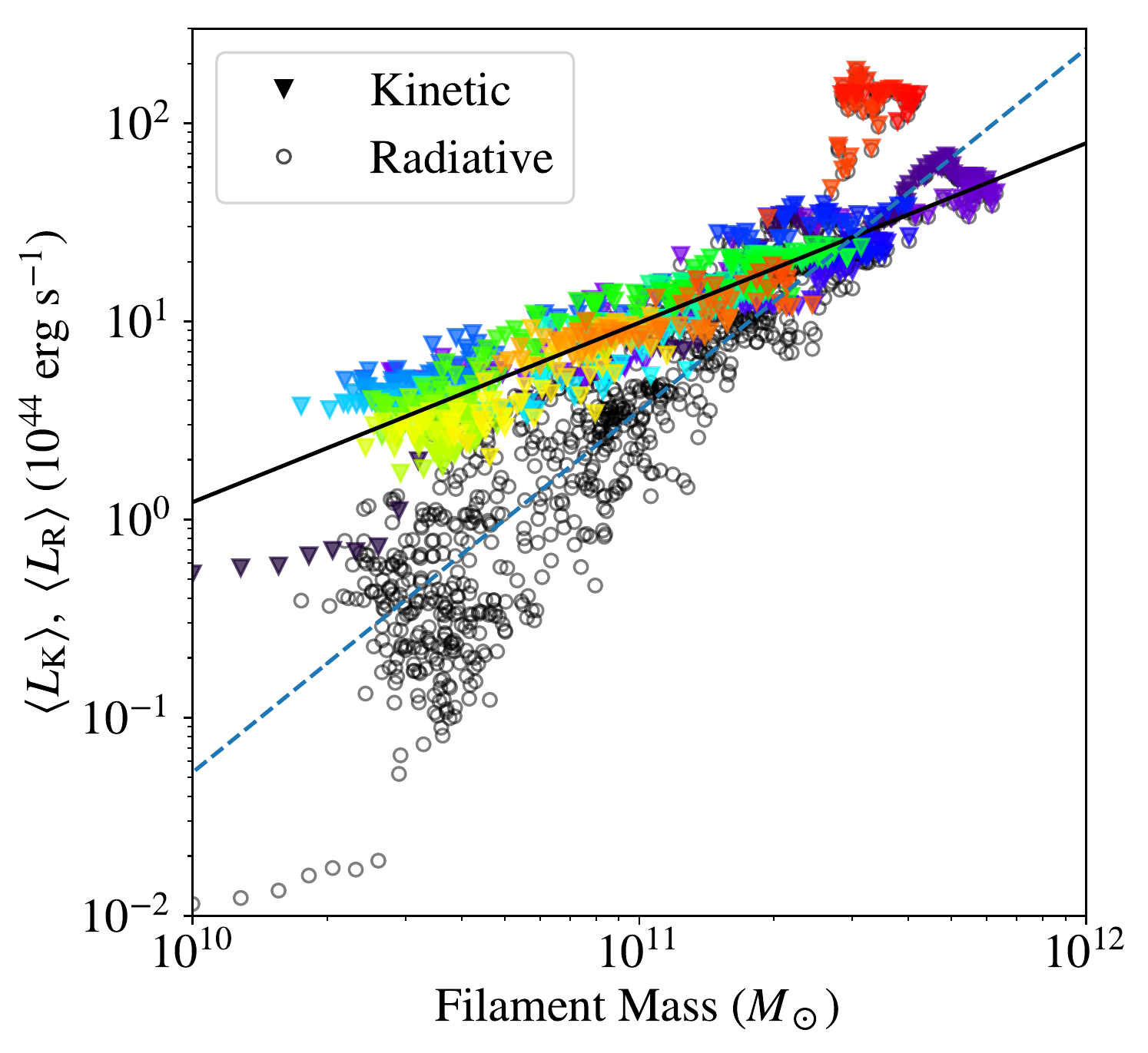}
\caption{{\it Left:} Evolution of the average kinetic (blue triangles), radiative luminosity (black circles), and H$\alpha$ filament mass (multicolor stars) measured in simulation RT02. {\it Right:} The kinetic (multicolor triangle) and radiative (black circle) luminosity as a function of the filament mass. The color marks the time in the simulation. Correlation of $\langle L_{\rm K}\rangle$ ($\langle L_{\rm R}\rangle$) with the H$\alpha$ filament mass for the first 9\,Gyr is represented by the solid black (dashed blue) line.}
\label{fig:f_mass}
\end{figure*}

The right panel of of Figure~\ref{fig:f_extent} also illustrates the properties of several observed clusters that host extended H$\alpha$ filaments \citep{McDonald2010, McDonald2015}. The kinetic luminosities for these systems are inferred from the size of their X-ray cavities \citep{Rafferty2006, McDonald2015}. It is interesting to note that in the context of our simulations the Phoenix cluster corresponds to systems in which a large amount of cold gas obstructs the expansion of the outflowing filaments. Our simulations suggest that the Perseus cluster should also be in this category, because its H$\alpha$ filaments extend in all directions from the cluster center \citep{Conselice2001}. The implication is that about $10^8\,$yr ago Perseus had a powerful AGN outburst with a luminosity of $\sim10^{45}\,{\rm erg\,s}^{-1}$. If so, the kinetic luminosity inferred from the size of the X-ray cavities in Perseus falls short of this value by about one order of magnitude. This apparent discrepancy may be explained if the central AGN has created multiple cavities during this feedback episode, some of which are not identified in observations.

\subsection{Mass of the Filaments} \label{sec:fila_mass}

\begin{figure*}[t!]
\includegraphics[height=0.29\linewidth]{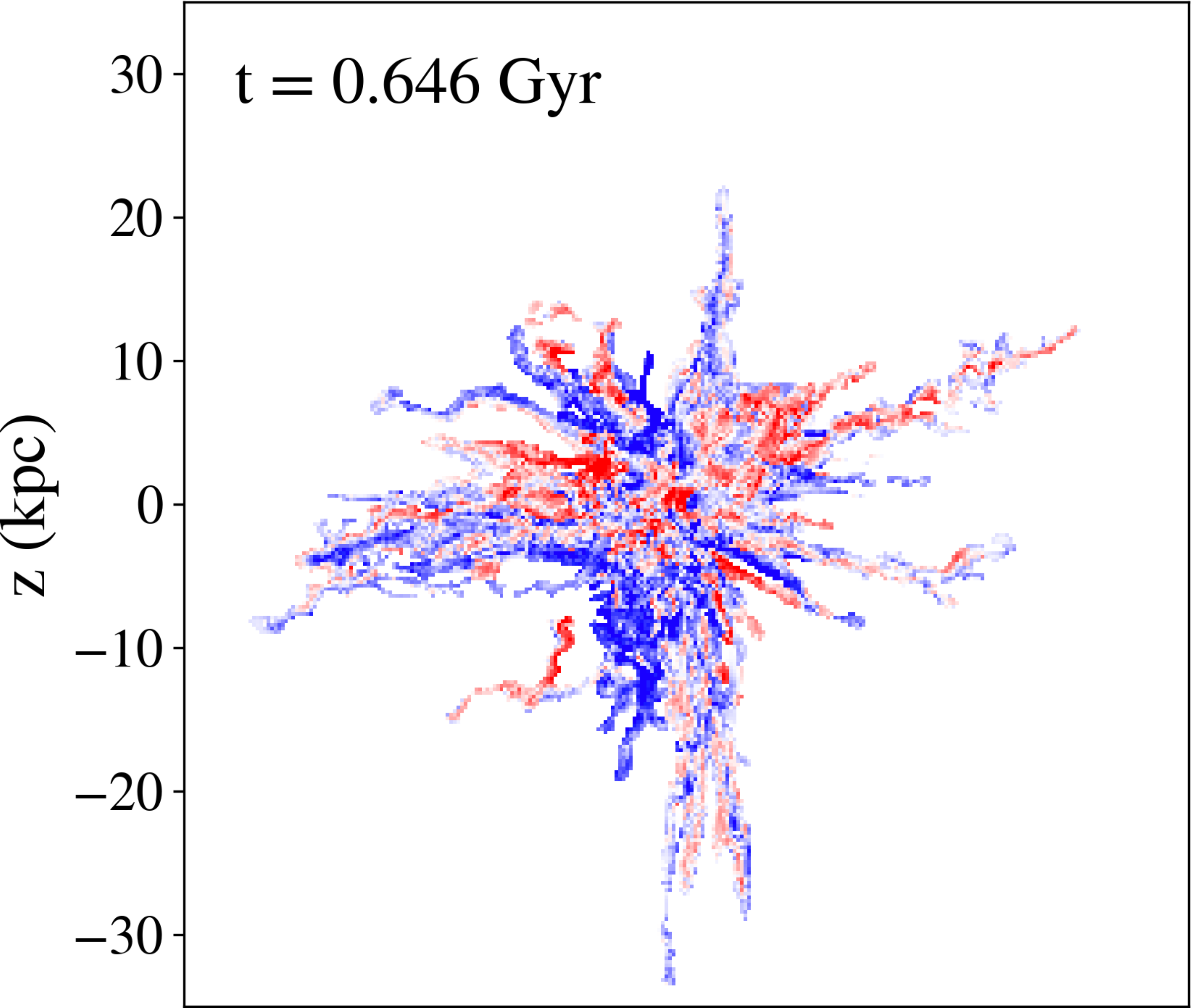}\includegraphics[height=0.29\linewidth]{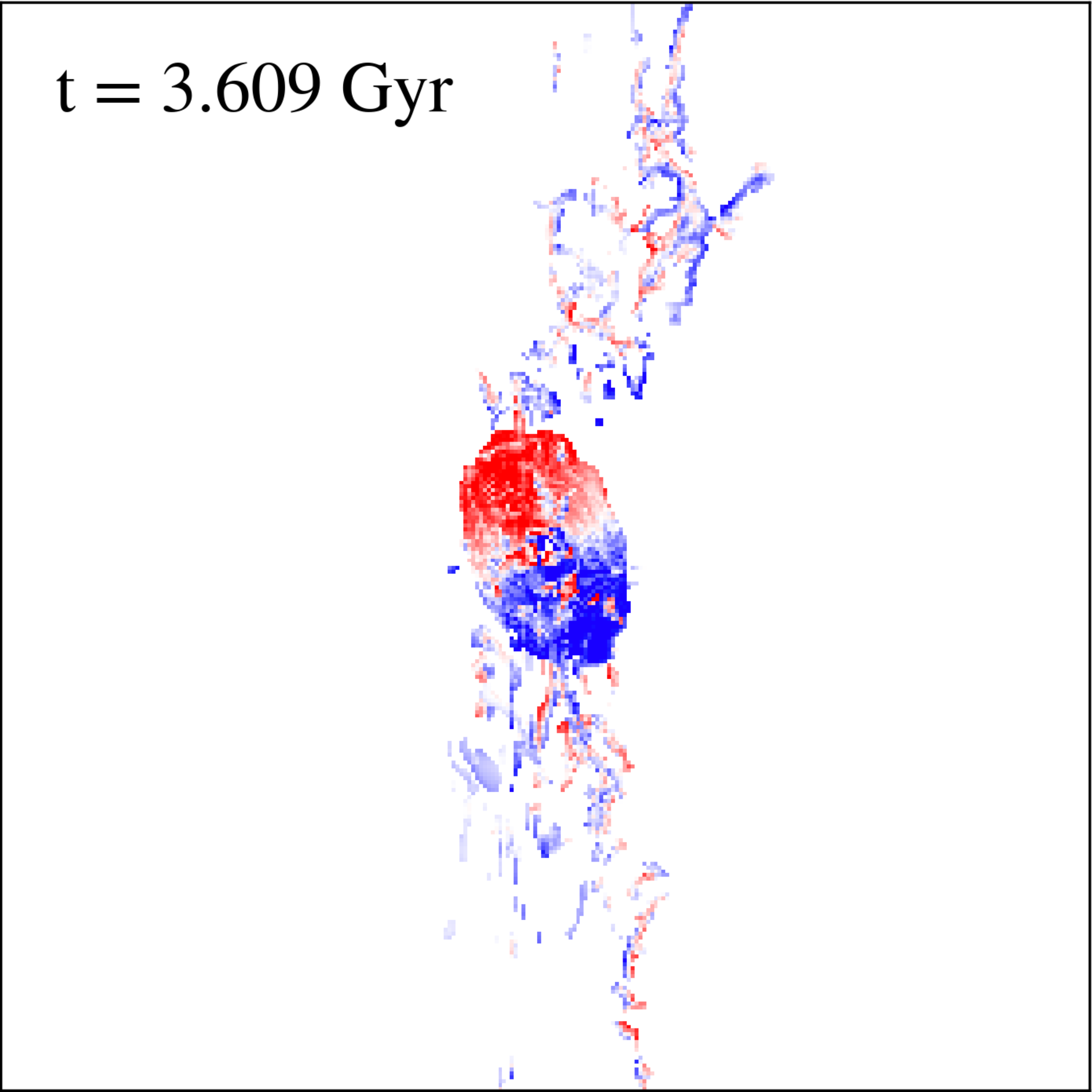}\includegraphics[height=0.29\linewidth]{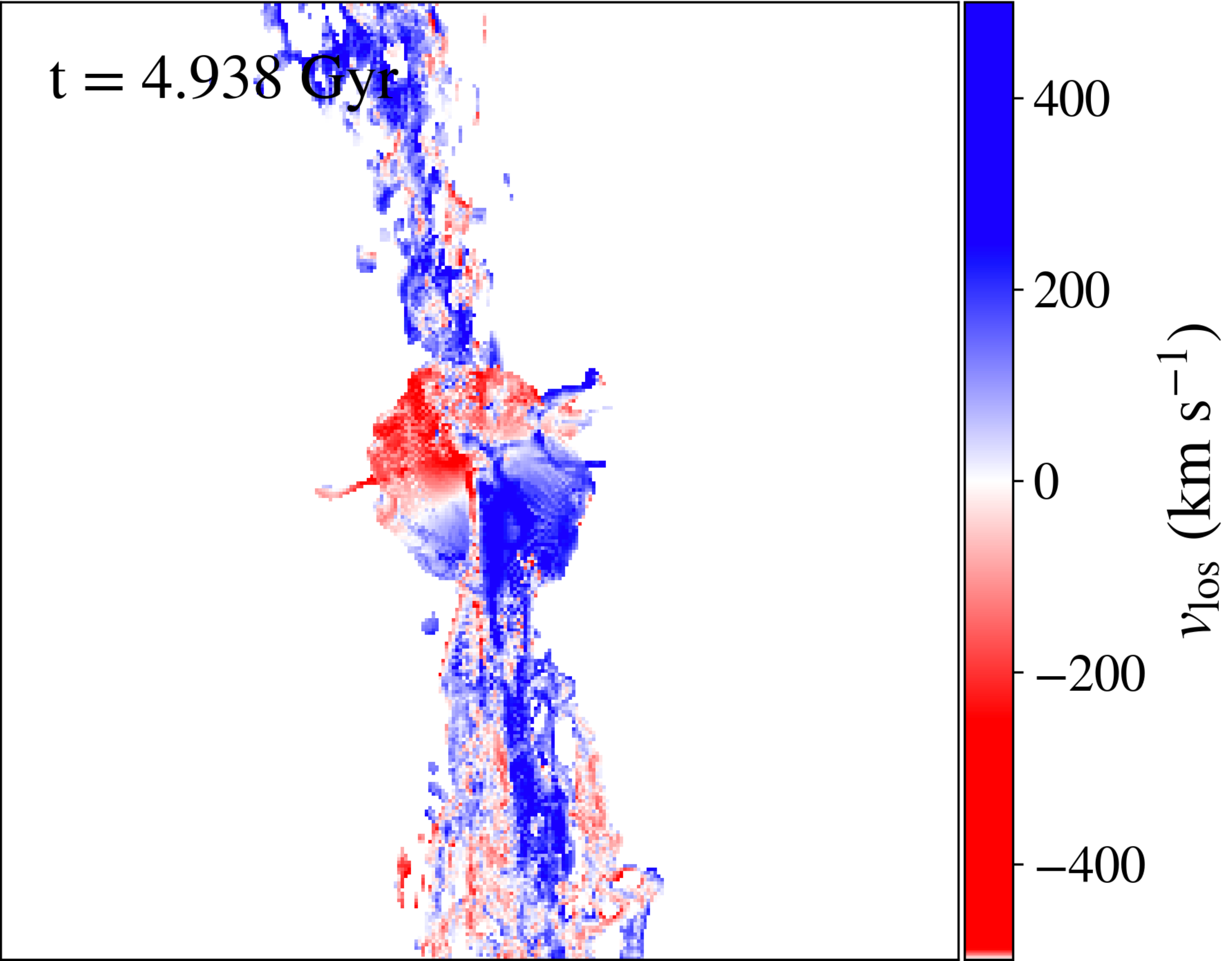}
\includegraphics[height=0.29\linewidth]{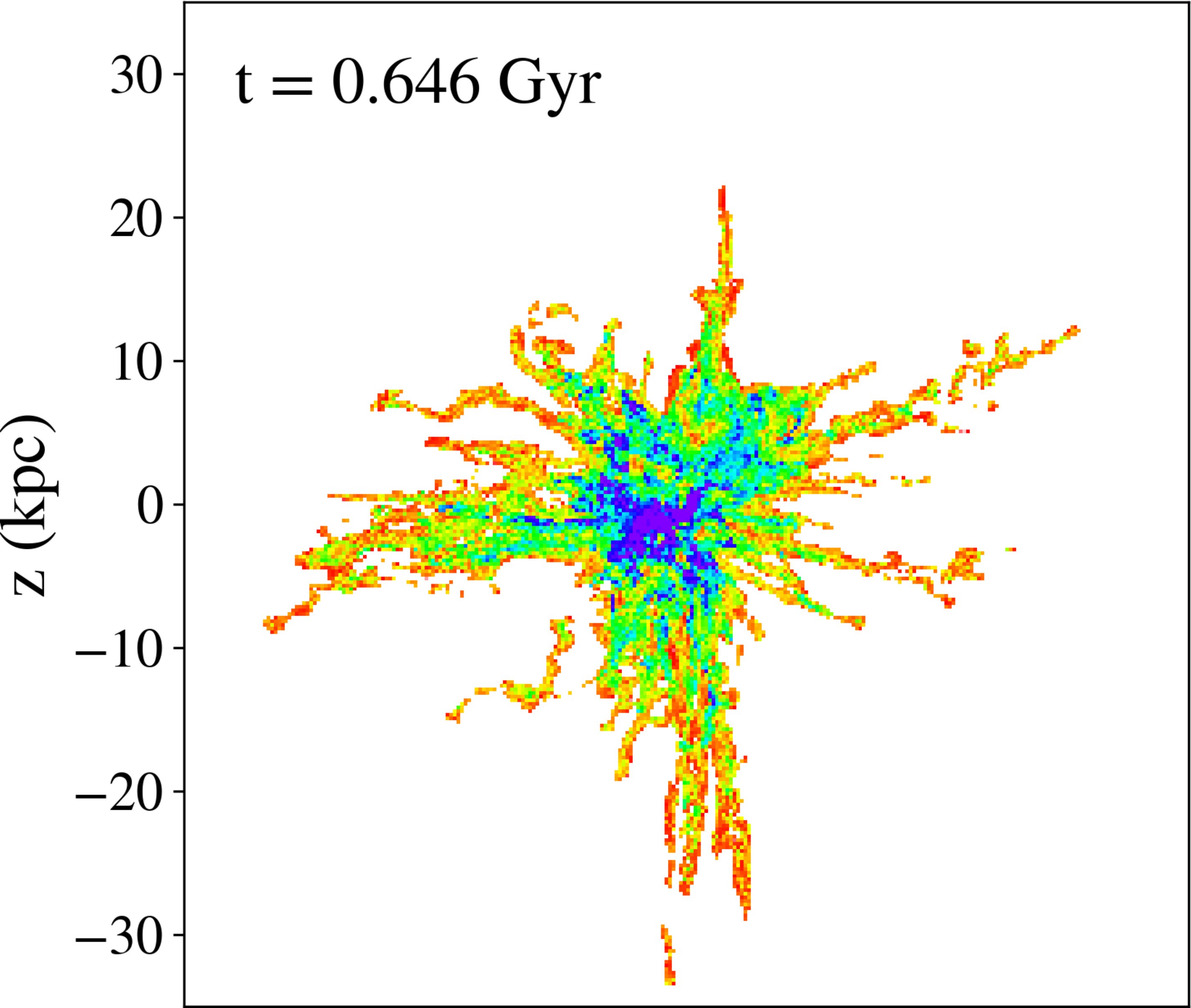}\includegraphics[height=0.29\linewidth]{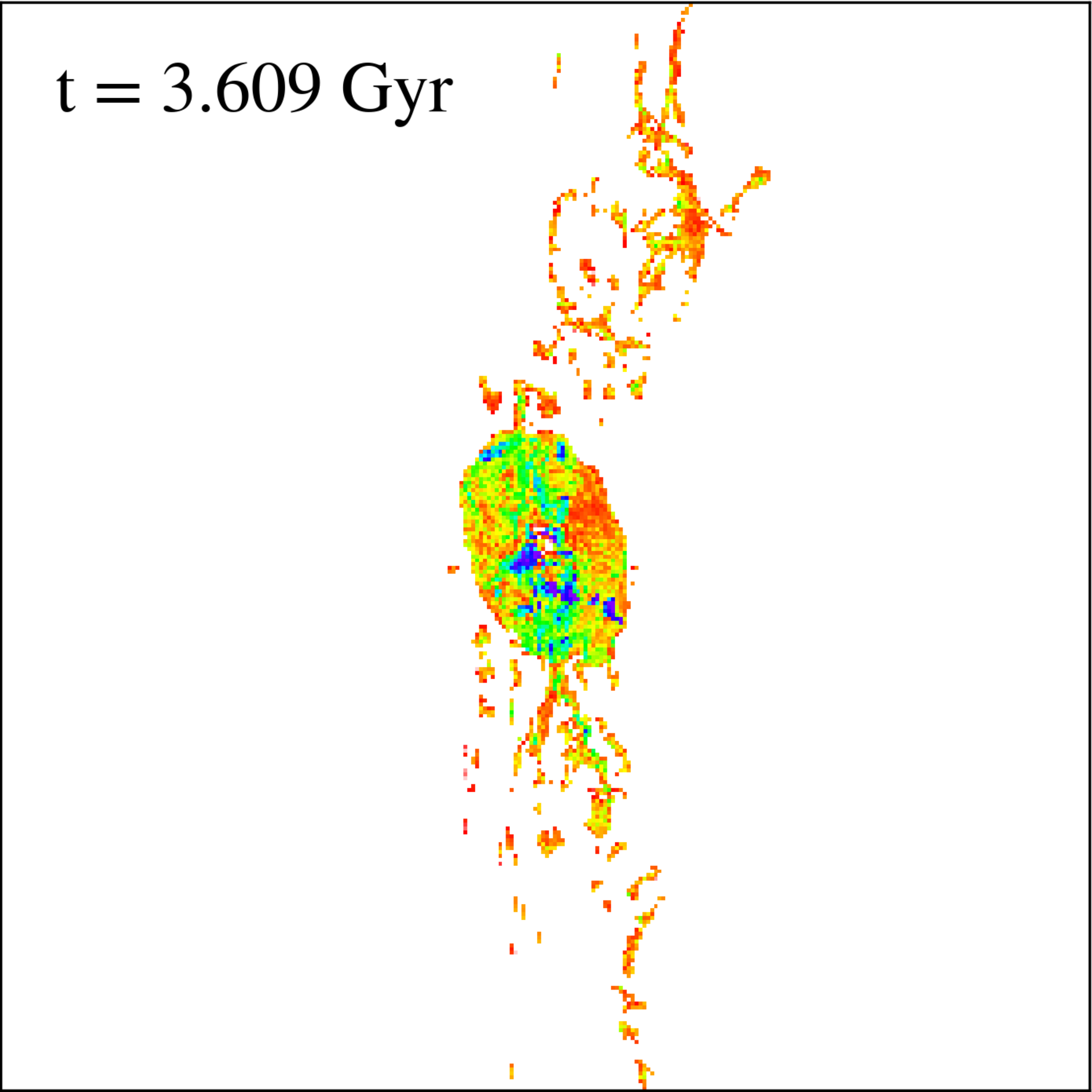}\includegraphics[height=0.29\linewidth]{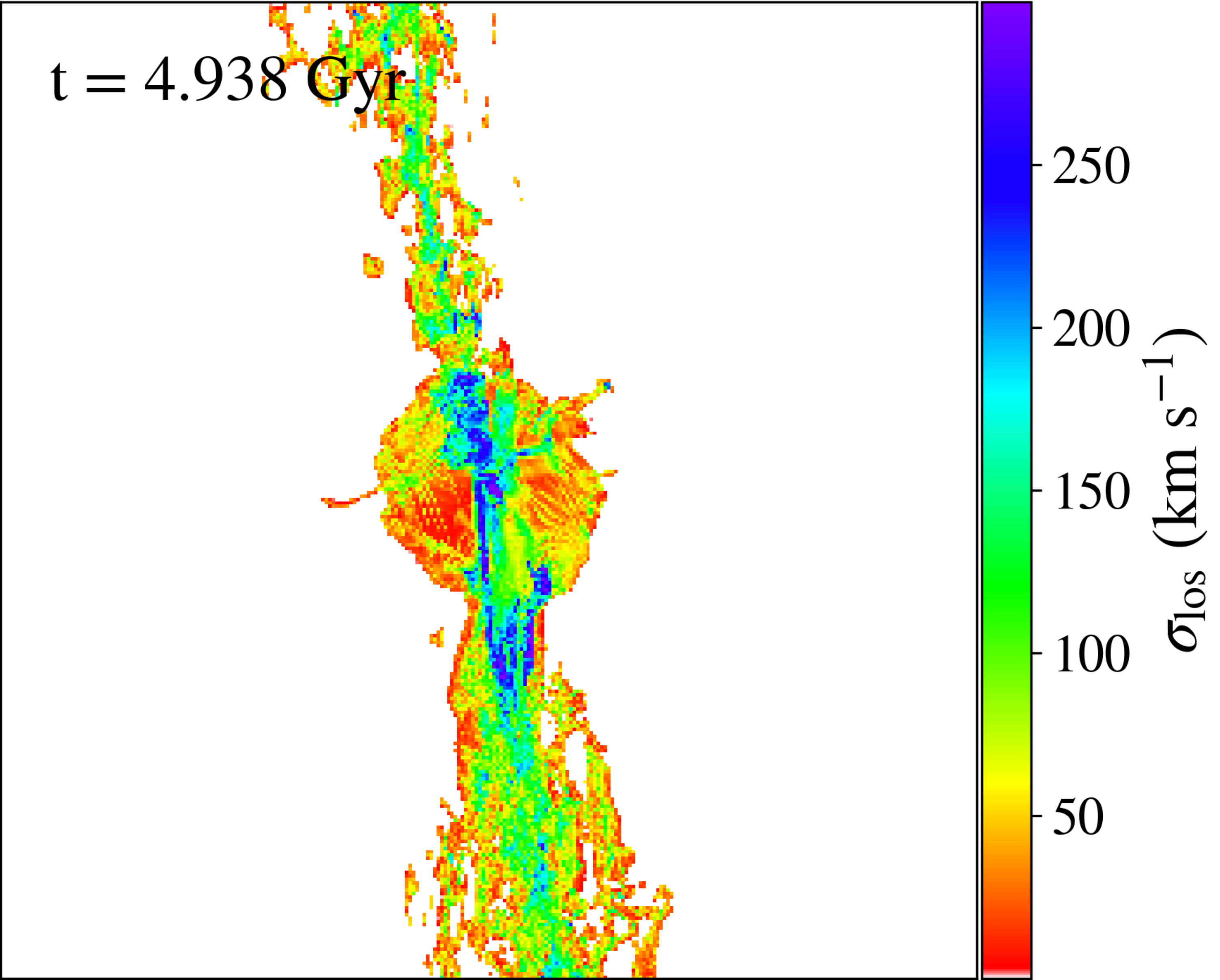}
\caption{The line-of-sight velocity (top) and velocity dispersion (bottom) maps calculated for the H$\alpha$ filaments that are predominantly radial ($t=0.65$\,Gyr) or collimated along the jet axis ($t=3.61$ and 4.94\,Gyr) in simulation RT02.}
\label{fig:f_vel}
\end{figure*}

In this section we investigate the degree of correlation between the AGN feedback luminosity and the mass of the H$\alpha$ filaments, $M_{\rm fil}$. The left panel of Figure~\ref{fig:f_mass} shows the evolution of the average AGN luminosity (both kinetic and radiative) and $M_{\rm fil}$ measured from simulation RT02. $\langle L_{\rm K}\rangle$ and $\langle L_{\rm R}\rangle$ are averaged over the dynamical timescale of the filaments, as described in Section~\ref{sec:fila_extent}. We measure $M_{\rm fil}$ as the mass of the $\textsc{H\,i}$ gas with rotational velocity $<300\,{\rm km\,s}^{-1}$. This criterion allows us to distinguish the filaments from a kinematically separate, rotationally supported cold gas disk that forms at the center of the cluster, and is visible in the middle and right panels of Figure~\ref{fig:f_vel}. 

The resulting relation between $\langle L_{\rm K}\rangle$ and $M_{\rm fil}$ is shown in the right panel of Figure~\ref{fig:f_mass}. In this case, a linear regression fit provides a good representation of the data in the first 9\,Gyr
\begin{equation}
\langle L_{\rm K}\rangle=(1.22\pm 0.03)\times 10^{44}\,{\rm erg\,s}^{-1}\left(\frac{M_{\rm fil}}{10^{10} M_\sun}\right)^{0.91\pm0.01}.
\label{eq3}
\end{equation}
In the context of our model, this implies $M_{\rm fil}^{0.9} \propto L_{\rm K} \propto \dot{M}_{\rm BH}$. 
If AGN feedback in real systems operates in the same way, this suggests that the mass (or equivalently luminosity) of the H$\alpha$ filaments can be used as a probe of the SMBH accretion rate. Similarly, the relation between $\langle L_{\rm R}\rangle$ and $M_{\rm fil}$ can be characterized by a power-law index of $1.83\pm0.03$, so that $L_{\rm R} \propto M_{\rm fil}^{1.8} \propto \dot{M}_{\rm BH}^2$.
This relationship of $L_{\rm K}$ and $L_{\rm R}$ with the SMBH mass accretion rate is a consequence of the prescription used in our simulations to describe the allocation of the AGN luminosity between the kinetic and radiative feedback in the radiatively inefficient accretion state (see Section~\ref{sec:sim}). Thus, the correlation with $M_{\rm fil}$ in Equation~\ref{eq3} 
applies to radiatively inefficient AGNs and may differ for radiatively efficient AGNs.

This point is illustrated by the high luminosity outburst at the end of the simulation, represented by the red data points, which do not fall on the correlation. These data points are associated with the radiatively efficient (high luminosity) state of the SMBH, when it appears as a radio-loud quasar. Large amounts of energy injected into the ICM by jets and radiation partially suppress the formation of the H$\alpha$ filaments, thus shifting the data points to the left of the relation. It is worth noting that both the AGN luminosity and H$\alpha$ filament mass are affected by the numerical resolution in RT runs in a similar way (both are higher in lower resolution runs, see a resolution study in Q18). Consequently, the correlation between the AGN luminosity and filament mass is maintained regardless of resolution.

\subsection{Velocity Distribution of the Filaments} \label{sec:f_vel}

If perturbation and turbulence associated with the AGN feedback catalyze the formation of cold gas in the ICM as hypothesized, one would expect that AGN luminosity should also correlate with the velocity dispersion of the H$\alpha$ filaments. We examine this hypothesis here. 

The top panels of Figure~\ref{fig:f_vel} illustrate the line-of-sight velocity ($v_\text{los}$) distribution of filaments measured from simulation RT02 at times $t=0.65, 3.61$ and 4.94\,Gyr. As discussed in Section~\ref{sec:fila_extent}, the early stages are characterized by the H$\alpha$ filaments that extend radially out in all directions. In later stages, however, the filaments are collimated along the jet axis (corresponding to the $z$-axis), thereby establishing a preferential axis that breaks the spherical symmetry in the velocity dispersion. The filament velocity in these images is weighted by the H$\alpha$ luminosity, in order to produce a map comparable to those obtained from observations. For this purpose, the H$\alpha$ luminosity of the gas is assumed to be proportional to the rate of recombination
\begin{equation}
L_{{\rm H}\alpha}\propto n_p n_e T_4^{-0.942-0.031\,\ln T_4} \,\,, 
\label{eq5}
\end{equation}
where $n_p$ and $n_e$ are the proton and electron number densities, and $T_4\equiv T/10^4\,{\rm K}$ \citep{Dong2011,Draine2011}. This emission is mainly contributed by the plasma with temperature $\sim10^{4}\,{\rm K}$, and we do not account for the attenuation of the H$\alpha$ photons due to absorption and scattering. 

The bottom panels of Figure~\ref{fig:f_vel} show the line-of-sight velocity dispersion ($\sigma_{\rm los}$) of the filaments in the same simulation snapshots. $\sigma_{\rm los}$ is calculated in each pixel of the image as a statistical dispersion around the average velocity in each pencil beam (with a cross section of $0.25\,{\rm kpc}\times0.25\,{\rm kpc}$) along the line of sight, weighted by the local H$\alpha$ luminosity. The resulting velocity maps have a lot of similarities with the H$\alpha$ filaments in Perseus and other clusters \citep{McDonald2012, Gendron-Marsolais2018}, as well as with other simulations of CCCs \citep{Li2018,Gaspari2018}.

\begin{figure}[t!]
\centering
\includegraphics[width=0.95\linewidth]{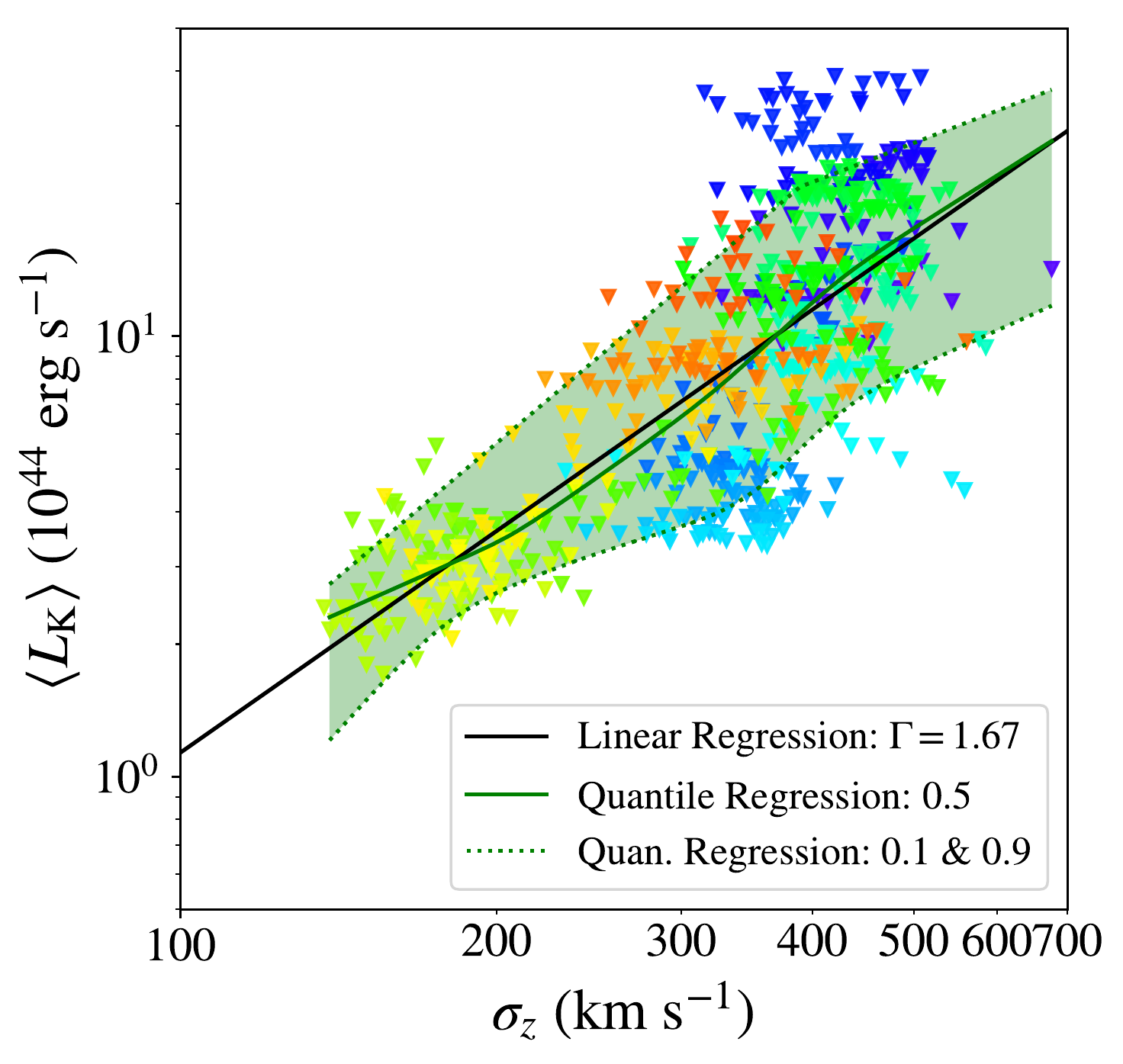}
\caption{$\langle L_{\rm K}\rangle$ as a function of the velocity dispersion of the H$\alpha$ filaments along the jet axis, $\sigma_z$. Data points correspond to $2<t<9$\,Gyr in simulation RT02, with the same color scheme as in Figure~\ref{fig:f_extent}.}
\label{fig:f_sigma}
\end{figure}

In order to investigate the relationship with the AGN feedback power, we calculate the velocity dispersion of the entire H$\alpha$ filamentary network in directions along and perpendicular to the jet axis. We find no statistically significant correlation of the perpendicular components with the feedback power, and measure their time averaged values to be $\langle\sigma_x\rangle=161\pm23\,{\rm km\,s}^{-1}$ and $\langle\sigma_y\rangle=162\pm33\,{\rm km\,s}^{-1}$, shown with one standard deviation. Similarly, during the first $2\,{\rm Gyr}$, we find no correlation with the value of velocity dispersion along the jet axis, $\langle\sigma_z\rangle=348\pm79\,{\rm km\,s}^{-1}$.  In the later stages ($2\,{\rm Gyr}<t<9\,{\rm Gyr}$), however, $\sigma_z$ correlates positively with the kinetic luminosity. This correlation is illustrated in Figure~\ref{fig:f_sigma} with an approximate power-law fit
\begin{equation}
\langle L_{\rm K}\rangle=(7.1\pm0.1)\times 10^{44}\,{\rm erg\,s}^{-1}\left(\frac{\sigma_z}{{300\,\rm km\,s}^{-1}}\right)^{1.67\pm0.05},
\end{equation}
and representative additive regression quantiles.
As a consequence of the assumptions in our feedback model, a corresponding relation exists between $\langle L_{\rm R}\rangle$ and $\sigma_z$. We do not show it here because the kinetic feedback is the primary cause of the correlation. If these measurements are applicable to the real H$\alpha$ filament systems in observed CCCs, they imply that the observed velocity dispersion is a function of the viewing angle, and may be used as an independent constraint on the spatial orientation of the jets.

\section{Discussion and Conclusions} \label{sec:conclusion}

In this Letter we use 3D radiation-hydrodynamic simulations of AGN feedback in galaxy clusters to examine and quantify the relationship between the observable properties of the H$\alpha$ filaments and the kinetic and radiative feedback from SMBHs in BCGs. We summarize the main findings and simplifying assumptions below. 

\begin{itemize}
\item We find that the AGN feedback in CCCs promotes the formation of spatially extended H$\alpha$ filaments and that their presence coincides with an ongoing outburst of AGN feedback. We identify two distinct distributions of filaments in our simulations. One is predominantly radial (similar to the Perseus cluster) and it arises when the expanding filaments collide and are scattered in all directions by the infalling filaments. The other exhibits collimated filaments along the jet axis and arises when the outflowing filaments can expand freely into the ICM.

\item The spatial extent of the collimated H$\alpha$ filaments correlates with the kinetic luminosity of AGN feedback in such a way that more luminous AGNs tend to produce more extended H$\alpha$ filament networks. This correlation can be used as a diagnostic for AGN feedback in nearby clusters, in which the extent and distribution of the filaments can be spatially resolved. 
\item We also report a correlation between the kinetic luminosity of AGN feedback and the H$\alpha$ filament mass. The correlation implies that $M_{\rm fil}^{0.9} \propto L_{\rm K} \propto \dot{M}_{\rm BH}$ and therefore the filament mass can be used to probe the AGN feedback and the rate of fueling of the central SMBH. This correlation can in principle be applied to about half of all clusters, which exhibit nuclear H$\alpha$ emission \citep{McDonald2010}, regardless of whether or not their filaments are spatially resolved. 
\item We find a correlation between the kinetic luminosity of AGN feedback and velocity dispersion along the jet axis ($\sigma_z$) only in the case when the filaments are collimated along the jet axis and expanding freely into the ICM. We find no statistically significant correlation with the velocity dispersion measured perpendicular to the jet axis or with velocity dispersion of the filaments with radial distribution. If applicable to the H$\alpha$ filament systems in observed CCCs, this correlation may be used as an independent constraint on the spatial orientation of the jets.
\end{itemize}

Our analysis assumes that the H$\alpha$ photons are produced by the recombination of thermal plasma, aided by AGN feedback.  We do not consider the effect of additional sources of ionizing radiation (like star forming regions), cosmic rays, or magnetic fields on the properties of the H$\alpha$ filaments or their emissivity. We do not explicitly model the chemistry of molecular gas or star formation in our simulations, and are thus unable to make quantitative statements about them. 


\acknowledgments
Support for this work was provided by the National Aeronautics and Space Administration through Chandra Award Number TM7-18008X issued by the Chandra X-ray Center, which is operated by the Smithsonian Astrophysical Observatory for and on behalf of the National Aeronautics Space Administration under contract NAS803060. TB acknowledges support from the National Science Foundation under grant No. NSF AST-1333360 during the early stages of this work. 



\end{CJK}
\end{document}